# The Role of Mobile and Social Media Services in Enhancing Freedom of Expression: Opportunities, Challenges, and Prospects for Local Platform Development in Uganda's Digital Ecosystem

**Bazigu Alex** *

Department of Networks, College of Computing and Information Sciences

Makerere University Kampala, Uganda

Email: bazigu.alex@gmail.com

Faculty of Science and Technology

Victoria University, Kampala, Uganda

Email: abazigu@vu.ac.ug

*Corresponding Author

**Jacinta Nafuna** *

Department of Networks, College of Computing and Information Sciences

Makerere University, Kampala, Uganda

Email: jacintanafuna51@gmail.com

*Corresponding Author

**Dr. Drake Mirembe (PhD)**

Department of Networks, College of Computing and Information Sciences

Makerere University, Kampala, Uganda

drake.mirembe@mak.ac.ug

**Abstract:**

Utilizing mobile and social media platforms is a transformative approach to enhancing freedom of expression and fostering digital engagement. However, Uganda's digital ecosystem faces challenges such as restrictive legislation, financial barriers, and the absence of localized platforms tailored to cultural contexts. This study employed a mixed-methods approach to explore how these platforms influence public discourse, activism, and civic participation while highlighting opportunities for local innovation. The research further identified the critical need for regulatory reforms, investments in digital literacy, and collaborative efforts to develop sustainable and culturally relevant platforms, ensuring a more inclusive and empowered digital society.

**Keywords:** Freedom of Expression, Mobile Services, Social Media Platforms, Local Digital Innovation, Uganda's Digital Ecosystem

# 1    Introduction

The evolution of mobile services has significantly influenced global communication patterns, transforming them from simple voice and text functionalities to powerful platforms for connectivity, commerce, and social mobilization (Guarda et al., 2020; Otto and Kruikemeier, 2023a). Early mobile communication technologies, such as SMS, provided basic yet critical channels for exchanging information, particularly in regions with limited access to traditional communication infrastructure (Wraycastle, 2024). In Africa, for instance, SMS-based services like M-Pesa revolutionized financial inclusion and showcased the potential of mobile technologies to address local needs (Jack and Suri, 2011).

With the advent of mobile internet and smartphones, social media platforms emerged as dominant communication tools (Otto & Kruikemeier, 2023b; Wang et al., 2019). Platforms like Facebook, Twitter, and WhatsApp extended the scope of mobile communication, enabling users to share multimedia content, organize events, and engage in political discourse. These platforms' accessibility and immediacy made them indispensable during key global events, such as the Arab Spring (2011), where mobile technologies played a pivotal role in mobilizing protests and bypassing state censorship (Howard and Hussain, 2013).

In an era where digital technologies shape public discourse, mobile and social media platforms have become essential for freedom of expression (Xie, 2024). These technologies have fundamentally transformed how individuals and societies interact, offering platforms for the exchange of ideas, mobilization for causes, and the circumvention of traditional restrictions on expression. Over the past two decades, the proliferation of smartphones and affordable internet access has enabled billions of people worldwide to participate in digital ecosystems, fostering unprecedented levels of connectivity (Poushter et al., 2018).

The relationship between mobile and social media services and freedom of expression in Uganda can be understood through three primary theoretical lenses. The Public Sphere Theory, developed by Habermas (1962), describes a space where individuals engage in discourse on societal issues and influence political outcomes (Habermas, 2010). Mobile and social media platforms serve as modern digital public spheres, allowing Ugandans to mobilize, organize protests, and challenge political repression. For example, during the 2021 elections, platforms like Facebook and Twitter enabled citizens to voice dissent and participate in political discourse (Muzee and Enaifoghe, 2020). However, critics argue that digital public spheres risk creating echo chambers, which limit exposure to diverse views and hinder meaningful debate (Sunstein, 2001).

The Technological Determinism suggests that advancements in technology drive societal change, shaping communication, power structures, and civic participation (Islas, 2016). Social media platforms like Twitter, Facebook, and TikTok have empowered Ugandans to circumvent traditional media restrictions, as seen in campaigns like #FreeBobiWine. While these technologies foster empowerment, they are also vulnerable to manipulation through misinformation and censorship, underscoring the need for critical governance of digital tools (Howard & Hussain, 2013).

The Freedom of Information Framework emphasizes access to information as fundamental to freedom of expression and active citizenship. Mobile and social media platforms have expanded access to information in Uganda, especially where traditional media is constrained by political pressures (CIPESA, 2023). However, barriers such as the OTT tax and internet shutdowns during critical moments restrict the platforms' full potential, reflecting ongoing tensions between state control and digital rights (Maractho, 2023).

Mobile communication has evolved drastically, with global platforms like Facebook, Twitter (now X), and WhatsApp dominating worldwide enabling instant communication and mobilization while localized platforms such as WeChat in China and LINE in Japan and Taiwan, KakaoTalk in South Korea, Zalo in Vietnam, and Gojek in Indonesia demonstrate how platforms adapt to regional sociocultural, economic, and political contexts (van Dijck et al., 2018). Around the world, these social media tools have empowered individuals to exchange ideas, mobilize for social and political causes, and bypass traditional censorship (Ong and Toh, 2023).

Around the world, social media has proven pivotal in mobilizing movements and amplifying voices. During the 2019–2020 pro-democracy protests in Hong Kong, platforms such as Telegram and Twitter were instrumental in organizing large-scale demonstrations (Protests and Hanberg, 2020; Shek, 2020) while during the #EndSARS campaign in Nigeria (2020), social media played a pivotal role in garnering global attention to police brutality (CIPESA, 2023; Dambo et al., 2022; Uwalaka, 2024). The Arab Spring (2011), often regarded as a turning point for digital activism, showcased the transformative potential of online platforms in mobilizing political movements (Howard and Hussain, 2013). In a similar vein, the #BlackLivesMatter movement highlighted how social media platforms can serve as critical tools for amplifying voices, fostering global solidarity, and organizing large-scale protests against systemic racism and police brutality (Freelon et al., 2016)

Africa has witnessed a surge in social media activism addressing local issues and driving societal change. Campaigns such as #AfricaIsNotAShithole celebrate African identity, while others like #SaveRiverNyando in Kenya mobilized environmental advocacy. Similarly, #JusticeForKianjokomaBrothers sought justice for police brutality victims (Mwaura, 2021; Tinga et al., 2020; Twinomurinzi, 2024). In Uganda, campaigns like #FreeBobiWine during the 2021 elections and #ThisTaxMustGo opposing the OTT tax highlight the transformative role of social media in advocating for political freedoms and digital access (Boxell and Steinert-Threlkeld, 2022; Muzee and Enaifoghe, 2020). However, these movements face resistance through internet shutdowns and restrictive legislation, reflecting the tension between digital freedoms and state control (Balgobin and Dubus, 2022; Maractho, 2023).

While mobile and social media adoption in Uganda continues to grow, the country has yet to develop significant localized platforms, unlike other regions such as Asia. Localized platforms could address cultural and linguistic nuances while promoting data sovereignty. However, challenges such as inadequate funding, limited technological expertise, and restrictive regulatory frameworks hinder such innovation (Muzee and Enaifoghe, 2020). Furthermore, questions remain about whether locally developed platforms would gain traction in a market dominated by global giants.

1.2 Problem definition

In Uganda, government actions such as blocking Facebook during the 2021 elections (Amnesty, 2021), imposing the Over-the-Top (OTT) tax (Kisakye, 2021), and enforcing the Computer Misuse Act of 2011 have severely limited freedom of expression (Muhindo, 2022). These measures restrict access to information, hinder political discourse, and result in arrests for online content deemed critical of government figures. Despite over 13 million internet users, Uganda remains reliant on foreign-owned social media platforms, raising concerns about data privacy and sovereignty (Kemp, 2024). The lack of locally developed platforms further impairs these issues, stifling local innovation and economic growth. Without reform and investment in homegrown digital solutions, Uganda risks limiting its citizens' digital rights and missing opportunities for technological development.

1.3 Research Objectives

1. To analyze the impact of government regulations on freedom of expression in Uganda.
2. To explore the role of mobile and social media platforms in shaping political discourse and citizen engagement in Uganda.
3. To evaluate the challenges and opportunities for developing locally owned digital platforms that foster freedom of expression and digital innovation in Uganda.

**2 Material and methods**

This study adopted a mixed-methods design to investigate how mobile and social media services influence freedom of expression in Uganda, particularly focusing on government regulations and the potential for locally developed social media platforms. The central region of Uganda covering Kampala, Wakiso, and Mukono districts was selected for data collection due to its socio-economic and political diversity. These locations incorporate both urban and rural settings, ensuring broad representation of different demographic groups. Adult participants (18+ years old) who had used social media via mobile devices in the previous month were the primary focus of this study.

2.1 Research Approach

Between November,2024 and January, 2025, a combination of qualitative and quantitative approaches (Leavy, 2022) was used to secure both in-depth understanding and broader generalizability of the research. For the qualitative component, in-depth interviews and Focus Group Discussions (FGDs) were conducted to capture detailed insights into participants' opinions, experiences, and attitudes regarding social media regulation and usage. A purposive sampling strategy facilitated the selection of key informants namely the policymakers, legal experts, social media activists, and representatives of mobile service providers due to their direct engagement with Uganda's digital policies. FGDs were organized to accommodate groups of six (6) to eight (8) participants of varying ages, occupations, and digital access levels, recruited via community networks and local organizations. These sessions enabled participants to share perspectives on topics such as censorship, government interventions like the Over-the-Top tax (OTT), and the feasibility of locally hosted platforms.

2.2 Qualitative data collection

The quantitative data collection relied on a random sample of persons aged above 18 years meeting the inclusion criteria to complete a structured questionnaire. One hundred valid questionnaires were collected, thereby offering a dependable dataset for examining patterns of social media use, awareness of relevant regulations (such as the Computer Misuse Act), and willingness to adopt platforms developed within Uganda. In addition to primary data collection, a documentary review checklist provided supplementary qualitative information through the examination of documents such as internet penetration statistics, Uganda Human Rights Reports, the Computer Misuse Act, and historical data on mobile service development. This document review enriched the study's contextual background and supported triangulation of findings.

Face-to-face interviews were conducted for qualitative data collection, typically lasting 30–45 minutes, guided by semi-structured questions (Creswell and Creswell, 2022). Participants were first introduced to the researcher, briefed on the study's aim, and asked to sign a consent form. The interviews explored regulatory measures like social media shutdowns and taxes, personal experiences with censorship or self-censorship, and perceptions regarding the advantages or drawbacks of local platform development. FGDs, on the other hand, lasted roughly 45–60 minutes and were moderated using open-ended prompts to guide participants' discussions on social media usage, government oversight, and attitudes toward potential local innovations.

2.3 Quantitative data collection

For the quantitative data collection, a structured questionnaire measured respondent demographics, frequency of social media activity, and awareness of legislative directives (including the OTT tax). Likert-scale questions gathered data on perceptions of freedom of expression, trust in foreign-owned versus locally developed platforms, and how regulatory measures affected usage. This method allowed for efficient data collection from a larger population and the examination of relationships between key variables.

Following data collection, interviews and FGD recordings were transcribed verbatim, then imported into NVivo software for qualitative analysis. Content and thematic analyses were performed in line with Bougie and Sekaran (2019), enabling the researcher to organize and refine emergent themes and subthemes, such as "censorship experiences," "barriers to local innovation," and "data sovereignty." Irrelevant or redundant data were eliminated to maintain a clear focus on the study objectives (Bougie and Sekaran, 2019). This careful thematic sorting supported robust interpretation of participant viewpoints.

2.4 Data Analysis

Quantitative data from the structured questionnaires were coded, cleaned, and entered into the Statistical Package for Social Sciences (SPSS) version 26. The analysis proceeded through descriptive statistics (frequencies, percentages, means, and standard deviations) to offer an overview of social media usage trends, awareness of regulations, and respondent attitudes. Inferential tests (chi-square and Pearson correlation) were then used to investigate associations, such as whether heightened awareness of the OTT tax correlated with changes in social media usage behavior.

Throughout the research, strict ethical protocols were observed. Participants received thorough explanations of the study's purpose, gave informed consent, and could withdraw at any time without repercussions. Identifiers were replaced with pseudonyms to ensure confidentiality, and all data were kept secure to uphold participant privacy, sources and references such as published studies and official documents were properly acknowledged to avoid plagiarism, while the accuracy and integrity of all findings were carefully preserved.

Despite the study's rigorous design, several limitations arose. Some participants initially distrusted the researchers, suspecting motives beyond academic inquiry. Sensitivity around political matters and censorship occasionally led to self-censorship or cautious responses, and language barriers sometimes required translation, which risked minor translation inaccuracies. Nevertheless, the triangulation of methods (surveys, interviews, FGDs, and document reviews) helped to minimize biases and enhance the quality of the findings.

## 3. Results

This section presents the study's key findings, organized into distinct subsections. The results integrate both qualitative and quantitative data, emphasizing participant demographics, trends in social media usage, impact of regulatory measures, perceptions of freedom of expression, barriers to local platform development, opportunities for digital innovation, and the transformative role of digital activism.

3.1 Survey results.

The study included a sample of 101 participants from Uganda's central region, with 52.48% from Kampala, 27.72% from Wakiso, and 5.94% from Mukono, while 13.86% were from other districts. These regions represent varying levels of socio-political and digital engagement, ensuring a comprehensive analysis.

3.1.1 Participant Demographics

The demographic characteristics of the participants are summarized in Table 1 below.

Table 1. Participant Demographics by Gender, Age, Education, and District of Residence.

| **Bio-data** | **Category** | **Frequency** | **Valid Percentage** |
|---|---|---|---|
| Gender | Male | 71 | 70.3 |
|  | Female | 29 | 28.71 |
|  | Others | 1 | 0.99 |
| **Total** |  | **101** | **100.0** |
| Age Group | 18 to 34 years | 81 | 80.2 |
|  | 35 to 45 years | 15 | 14.85 |
|  | 46 to 60 years | 2 | 1.98 |
|  | Above 60 years | 3 | 2.97 |
| **Total** |  | **101** | **100.0** |
| Education | No Formal Education | 0 | 0.0 |
|  | Certificate | 10 | 9.9 |
|  | Diploma | 17 | 16.83 |
|  | Bachelor's Degree | 60 | 59.41 |
|  | Master's Degree | 12 | 11.88 |
|  | Ph.D. | 0 | 0.0 |
|  | Others | 2 | 1.98 |
| **Total** |  | **101** | **100.0** |
| District of Residence | Kampala | 53 | 52.48 |
|  | Mukono | 6 | 5.94 |
|  | Wakiso | 28 | 27.72 |
|  | Others | 14 | 13.86 |
| **Total** |  | **101** | **100.0** |

The study sample consisted of 101 participants, with a majority of 71 males (70.3%), 29 females (28.71%), and 1 respondent identifying as "Others" (0.99%), indicating a slightly male-dominated sample. In terms of age, the majority of respondents (80.2%) were aged between 18 and 34 years, followed by 14.85% aged 35 to 45 years, 1.98% aged 46 to 60 years, and 2.97% above 60 years, reflecting significant participation by younger individuals. Regarding education, the participants were highly educated, with 59.41% holding a Bachelor's Degree, 16.83% a Diploma, and 11.88% a Master's Degree, while 9.9% had a Certificate level qualification, and 1.98% fell into the "Others" category. Notably, none of the respondents reported having no formal education or a Ph.D. This shows that 75.29% of respondents had attained a Bachelor's Degree or higher.

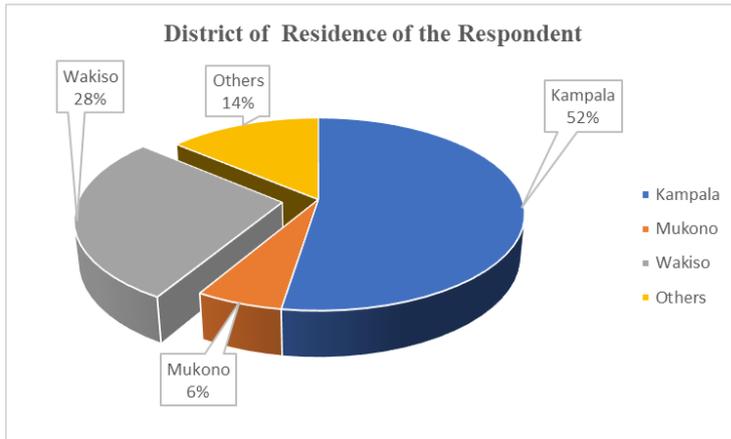

From Figure 1, the majority of participants resided in Kampala (52.48%) and Wakiso (27.72%), followed by Mukono (5.94%) and 13.86% from other districts such as Mbale, Mbarara, and Tororo. The sample's demographic profile highlights its urban-centred and well-educated nature, with a significant representation of youth, suggesting the respondents are likely to have high levels of engagement with mobile and social media platforms.

Figure 1. District of Residence of the Participants

### 3.1.2 Frequency of Social Media Use and Popular Social Media Platforms

The majority of respondents (93.07%) reported using social media multiple times a day, indicating heavy engagement. A smaller proportion (4.95%) use social media a few times a week, and only 1.98% reported using it once a day.

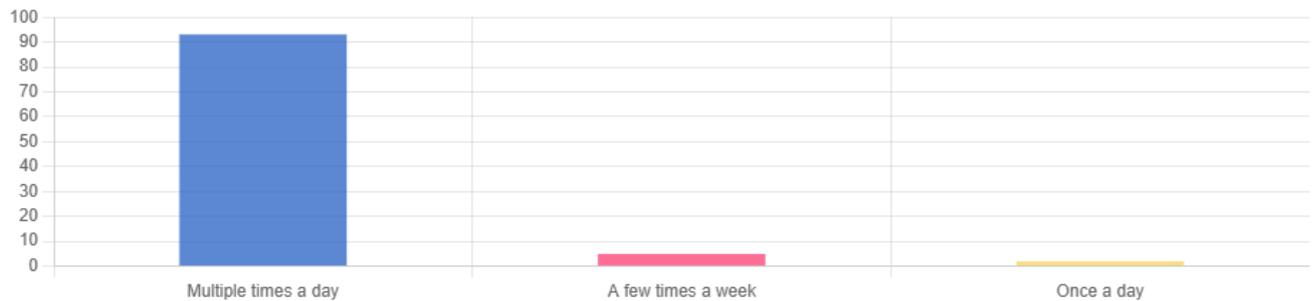

Figure 2. Frequency of Social Media Platform Usage

Platforms like YouTube (63.37%) and TikTok (62.38%) lead as the most popular platforms, followed by X (55.45%), Instagram (42.57%), and Facebook (37.62%), highlighting a strong preference for video and visual content among users.

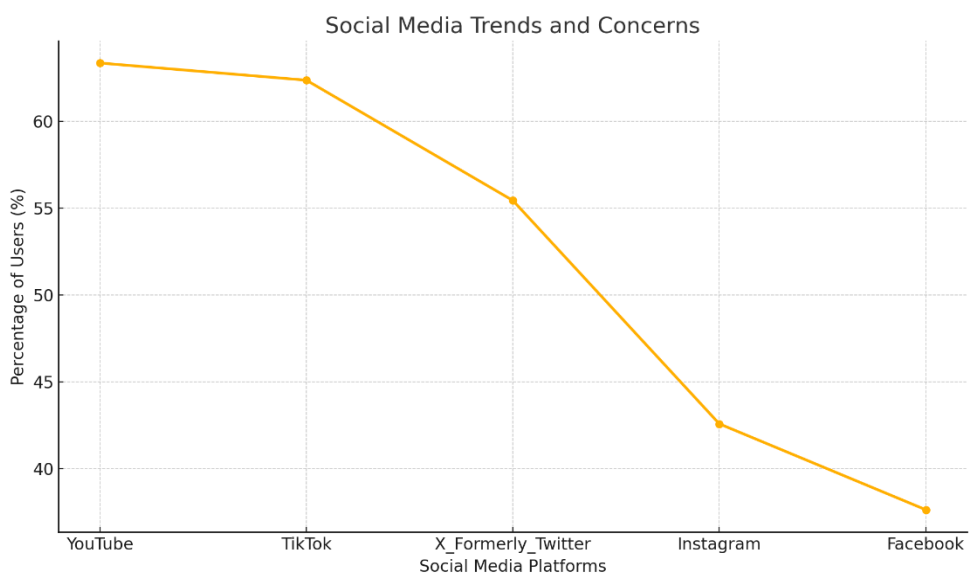

Figure 3. Social Media Trends and Concerns

Younger demographics (ages 18–34) demonstrated the highest levels of engagement with YouTube and TikTok, while older participants exhibited moderate but growing adoption of social media tools especially with Twitter and Instagram among the Female. Their social Media access is majorly through Smartphones at 97.03%, followed by desktops/laptops (48.51%) and tablets (11.88%), with minimal use of other devices (0.99%).

### 3.1.3 Impact of Regulations and Financial Constraints on Social Media Usage

Participants expressed significant concern over the government's regulatory practices, such as the 12% exercise duty tax on airtime and data and internet shutdowns on social media, A majority (65.35%) of respondents reported using social media more cautiously, while 50.5% avoid posting sensitive content online, reflecting heightened awareness and privacy concerns.

Financial constraints due to the tax led 24.75% to spend less time on social media, indicating a considerable barrier for users. Additionally, 20.79% resort to using VPNs or alternative methods to access platforms, and 15.84% feel their freedom of expression has been restricted while, 15.84% are unaware of how these regulations have affected them, suggesting a communication gap.

In FGDs, participants highlighted the 2021 election period as a turning point when regulatory measures severely limited online discourse, with many resorting to VPNs to maintain access. The imposition of these controls not only stifled freedom of expression but also amplified fears of surveillance and self-censorship, as shared by 62% of interviewees.

### 3.1.4 Perceptions of Freedom of Expression

The analysis shows that 60% of respondents believe social media enhances their ability to express opinions, highlighting its role in communication as illustrated in Figure 4. However, concerns about harassment remain unclear, emphasizing the need to balance freedom of expression with user safety in digital spaces.

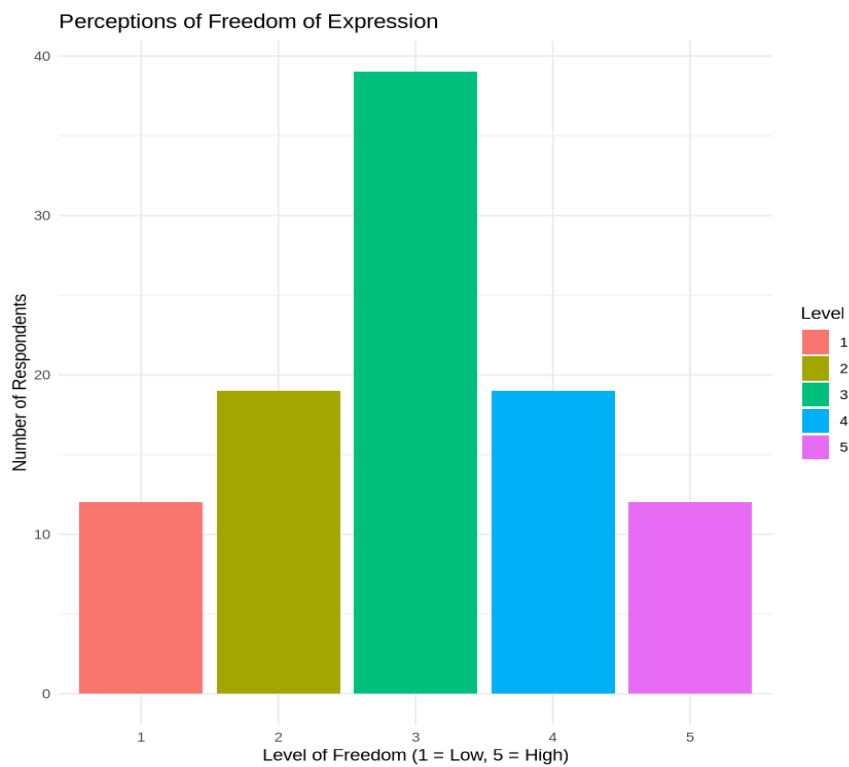

Figure 4. Perception of freedom of expression on the Social Media Platforms

Thematic analysis of qualitative responses revealed that while social media serve as a "digital public sphere," its potential was undermined by government censorship, misinformation, and lack of digital literacy among some users.

### 3.1.5 Barriers to Local Platform Development

Discussions in FGDs and interviews identified critical barriers to developing local platforms, including inadequate funding, insufficient technological expertise, and restrictive regulatory frameworks.

From the data, 42.57% of respondents expressed interest in using a locally developed social media platform, but only if it meets their specific needs. Another 38.61% indicated they would "definitely" use such a platform. Together, this highlights that more than 80% of respondents are open to local platforms, provided they meet expectations. Meanwhile, 27.72% are uncertain and prefer to decide based on features and usability. However, 10.89% prefer international platforms, citing trust or familiarity, and 5.94% explicitly do not trust locally developed platforms.

Table 2. Willingness to adopt locally developed Social Media Platform

| Value | Frequency | Percentage |
| --- | --- | --- |
| **Yes, but only if it meets my needs** | 43 | 42.57 |
| **Yes, definitely** | 39 | 38.61 |
| **Maybe, depending on its features and usability** | 28 | 27.72 |
| **No, I prefer international platforms** | 11 | 10.89 |
| **No, I don't trust locally developed platforms** | 6 | 5.94 |

During the FGD, participants unanimously agreed that localized platforms could address cultural and linguistic nuances while promoting data sovereignty. However, skepticism about their sustainability in a market dominated by global giants persisted. Quantitative findings corroborated these views, with only 22% of respondents expressing confidence in the success of Ugandan-developed platforms.

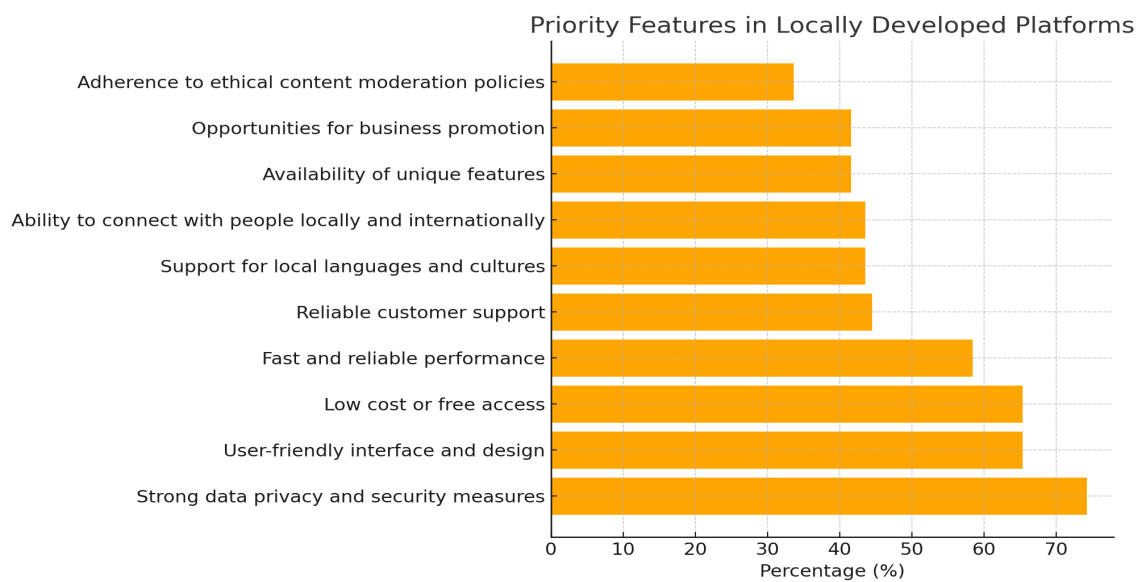

Figure 5. Priority Features required in a locally developed Social media platform.

### 3.1.6 Opportunities for Digital Innovation

Despite the challenges, participants identified opportunities for leveraging mobile and social media services to foster freedom of expression. Increased investment in digital literacy initiatives, partnerships with local innovators, and supportive policy reforms were seen as crucial steps. Data from interviews and questionnaires highlighted that 75% of respondents supported reducing regulatory barriers to encourage innovation and attract investment in the local tech sector.

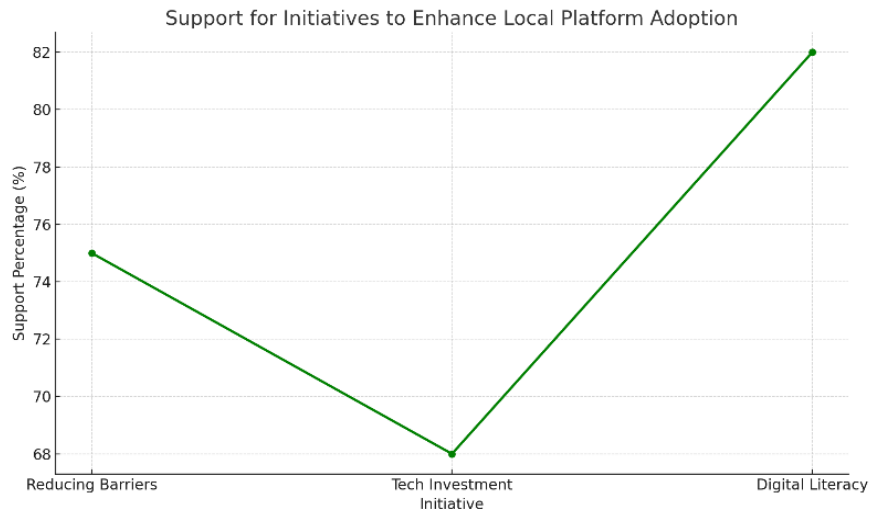

Figure 6. Support for Initiatives to Enhance Local Platform Adoption.

### 3.1.7 Digital Activism and Social Change

The analysis underscored the transformative potential of social media in driving social and political movements. Campaigns like #FreeBobiWine and #ThisTaxMustGo were cited as examples of successful digital activism in the interviewees conducted that amplified marginalized voices and mobilized collective action. However, FGDs noted that such efforts were often hindered by misinformation and inconsistent internet access, calling for improved media literacy and infrastructural support.

### 3.2 Statistical Analysis of Regulatory Awareness, Social Media Behavior, and Digital Freedom

Based on these descriptive trends, inferential analyses were conducted to evaluate the relationships between the variables. The results of the Chi-square and Pearson correlation tests are summarized below.

Table 3. Correlation between Awareness and Behavior Change

| Test | Value | p-value | Interpretation |
| --- | --- | --- | --- |
| **Chi-square Test ($\chi^2$)** | 67.73 | 0.977 | No significant association between awareness and social media behaviour ($p > 0.05$). |
| **Pearson Correlation (r)** | 0.132 | 0.193 | Weak and insignificant correlation ($p > 0.05$). |

From Based on these descriptive trends, inferential analyses were conducted to evaluate the relationships between the variables. The results of the Chi-square and Pearson correlation tests are summarized below.

Table 3, A Chi-square test ($\chi^2$ = 67.73, p = 0.977) revealed no significant association between awareness of regulatory measures and changes in social media usage behavior. Similarly, Pearson correlation analysis (r = 0.132, p = 0.193) indicated a weak and statistically insignificant relationship between perceptions of digital freedom and willingness to adopt local platforms. These findings suggest that neither regulatory awareness nor perceptions of digital freedom strongly influence user behavior or platform adoption in this context.

## 4. Discussion

Mobile and social media platforms in Uganda serve as both enablers and inhibitors of freedom of expression. While these tools empower citizens to engage in political discourse and mobilize social movements, challenges such as government restrictions, low digital literacy, and financial barriers hinder their full potential. The 12% exercise duty tax on airtime and data, and periodic internet shutdowns represent ongoing tensions between state control and digital freedoms. These measures disproportionately limit marginalized communities' access to digital platforms, fostering widespread concerns about censorship and privacy. Many users (65.35%) reported altering their social media behavior due to these pressures, reflecting the chilling effect of regulatory controls.

Uganda's reliance on global social media giants raises issues of data sovereignty and cultural relevance. Limited funding, technological expertise, and restrictive policies are key barriers to developing local platforms. Nonetheless, over 80% of respondents expressed willingness to adopt localized platforms if they meet security and usability standards, highlighting the potential for homegrown innovation.

Despite these challenges, significant opportunities exist to strengthen Uganda's digital ecosystem. Investments in digital literacy can empower citizens to use online platforms effectively, while partnerships among government, private sector, and academia can drive the creation of culturally relevant platforms. Reducing financial and infrastructural barriers through regulatory reforms can also promote broader inclusivity in digital access.

## 5. Conclusion and Recommendation

Mobile and social media platforms are vital tools for freedom of expression in Uganda. However, restrictive regulations, financial constraints, and limited local innovation hinder their full potential. Addressing these issues through targeted reforms and investments is essential for fostering a more inclusive digital ecosystem.

The government should remove the excise duty tax on data and airtime to lower financial barriers to internet access. Policies promoting the development of local platforms through funding and partnerships must be strengthened to encourage innovation and ensure cultural representation. The Ministry of ICT and National Guidance should enhance digital literacy by launching nationwide campaigns and integrating relevant modules into the education system. The expansion of broadband infrastructure in rural and underserved areas is crucial to reducing the digital divide. Collaboration between civil society and the government is essential to support open dialogue and advance digital activism.

Acknowledgment

The work was not funded by anyone. And the quantitative data can be accessed from the link